\documentclass{sig-alternate}
\usepackage[utf8]{inputenc}
\usepackage{graphicx}
\usepackage{cite}

\begin{document}




\title{SentiBubbles: Topic Modeling and Sentiment\\ Visualization of Entity-centric Tweets}

\numberofauthors{2} 
%
\author{
%
%
\alignauthor
Jo\~ao Oliveira, Mike Pinto\\
       \affaddr{DEI-FEUP}\\
       \affaddr{University of Porto}\\
       \affaddr{R Dr. Roberto Frias}\\
       \affaddr{Porto, Portugal}\\
       \email{pssc@fe.up.pt}\\
\and
\alignauthor
Pedro Saleiro, Jorge Teixeira\\
       \affaddr{DEI-FEUP, LIACC}\\
       \affaddr{University of Porto}\\
       \affaddr{R Dr. Roberto Frias}\\
       \affaddr{Porto, Portugal}\\
       \email{pssc@fe.up.pt}\\
}
\maketitle

\begin{abstract}
Social Media users tend to mention entities when reacting to news events. The main purpose of this work is to create entity-centric aggregations of tweets on a daily basis. By applying topic modeling and sentiment analysis, we create data visualization insights about current events and people reactions to those events from an entity-centric perspective. 
\end{abstract}


\section{Introduction}

Entities play a central role in the interplay between social media and online news \cite{Saleiro2016}. Everyday millions of tweets are generated about local and global news, including people reactions and opinions regarding the events displayed on those news stories. Trending personalities, organizations, companies or geographic locations are building blocks of news stories and their comments. We propose to extract entities from tweets and their associated context in order to understand what is being said on Twitter about those entities and consequently to create a picture of people reactions to recent events. 

With this in mind and using text mining techniques, this work explores and evaluates ways to characterize given entities by finding: (a) the main terms that define that entity and (b) the sentiment associated with it. To accomplish these goals we use topic modeling \cite{Blei:2003:LDA:944919.944937} to extract topics and relevant terms and phrases of daily entity-tweets aggregations, as well as, sentiment analysis \cite{Kim:2004:DSO:1220355.1220555}  to extract polarity of frequent subjective terms associated with the entities. Since public opinion is, in most cases, not constant through time, this analysis is performed on a daily basis. Finally we create a data visualization of topics and sentiment that aims to display these two dimensions in an unified and intelligible way. 

The combination of Topic Modeling and Sentiment Analysis has been attempted before: one example is a model called TSM - Topic-Sentiment Mixture Model \cite{Mei:2007:TSM:1242572.1242596} that can be applied to any Weblog to determine a correlation between topic and sentiment. Another similar model has been proposed proposed \cite{Lin:2009:JSM:1645953.1646003} in which the topic extraction is achieved using LDA, similarly to the model that will be presented. Our work distinguishes from previous work by relying on daily entity-centric aggregations of tweets to create a meta-document which will be used as input for topic modeling and sentiment analysis.

\begin{figure}[t!]
\centering
\includegraphics[width=0.7\columnwidth]{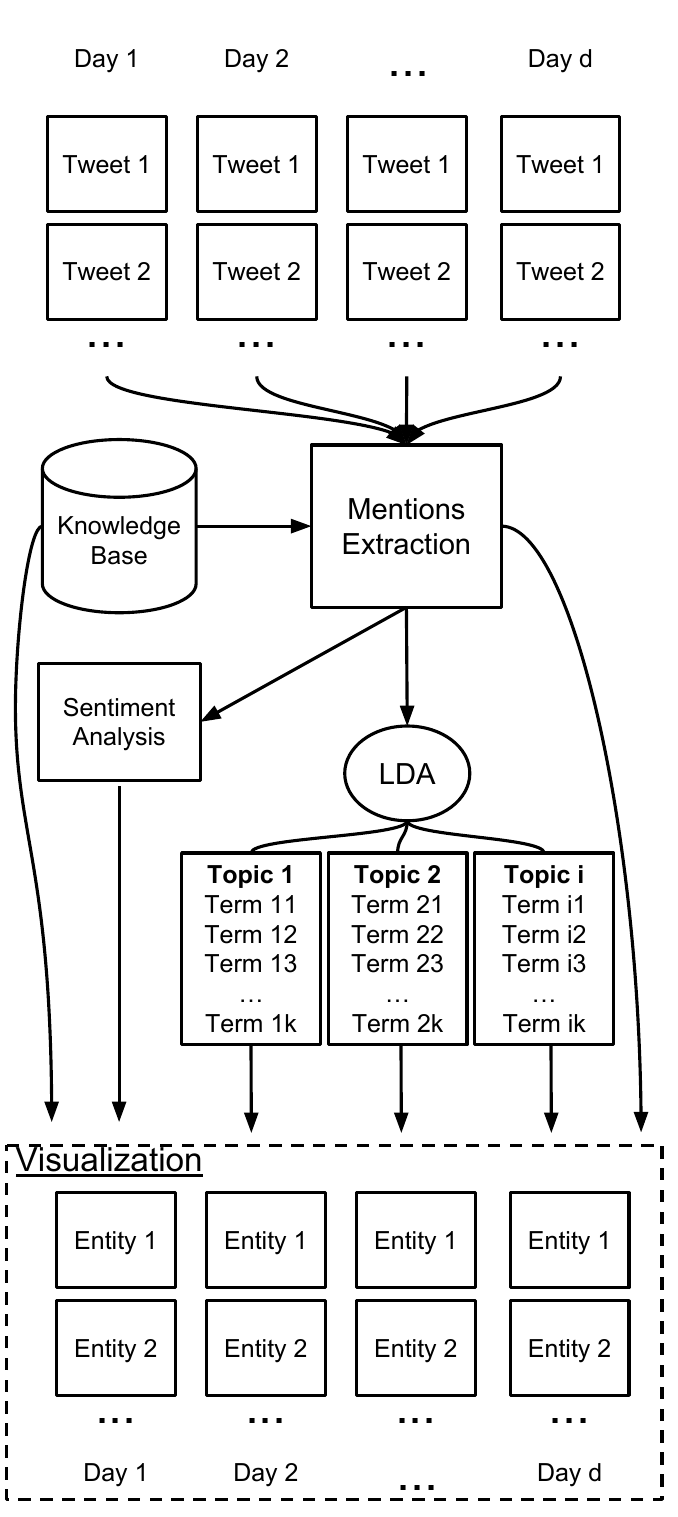}
\caption{The system pipeline}
\end{figure}

\section{Methodology}
The main goal of the proposed system is to obtain a characterization of a certain entity regarding both mentioned topics and sentiment throughout time, i.e. obtain a classification for each entity/day combination. 
\\

\subsection{Tweets Collection}
Figure 1 depicts an overview of the data mining process pipeline applied in this work. To collect and process raw Twitter data, we use an online reputation monitoring platform \cite{saleiro2015popmine} which can be used by researchers interested in tracking entities on the web. It collects tweets from a pre-defined sample of users and applies named entity disambiguation \cite{saleiro2013popstar}. In this particular scenario, we use tweets from January 2014 to December 2015. In order to extract tweets related to an entity, two main characteristics must be defined: its canonical name, that should clearly identify it (e.g. ``Cristiano Ronaldo") and a set of keywords that most likely refer to that particular entity when mentioned in a sentence (e.g.``Ronaldo", ``CR7"). Entity related data is provided from a knowledge base of Portuguese entities. These can then be used to retrieve tweets from that entity, by selecting the ones that contain one or more of these keywords. 

\subsection{Tweets Pre-processing}
Before actually analyzing the text in the tweets, we apply the following operations:
\begin{enumerate}
\itemsep0em 
\item If any tweet has less than 40 characters it is discarded. These tweets are considered too small to have any meaningful content;
\item Remove all hyperlinks and special characters and convert all alphabetic characters to lower case;
\item Keywords used to find a particular entity are removed from tweets associated to it. This is done because these words do not contribute to either topic or sentiment;
\item A set of portuguese and english stopwords are removed - these contain very common and not meaningful words such as ``the" or ``a";
\item Every word with less than three characters is removed, except some whitelisted words that can actually be meaningful (e.g. ``PSD" may refer to a portuguese political party);
\end{enumerate}

These steps serve the purpose of sanitizing and improving the text, as well as eliminating some words that may undermine the results of the remaining steps.
The remaining words are then stored, organized by entity and day, e.g. all of the words in tweets related to Cristiano Ronaldo on the 10th of July, 2015.

\subsection{Topic Modeling}

Topic extraction is achieved using LDA, \cite{Blei:2003:LDA:944919.944937} which can determine the topics in a set of documents (a corpus) and a document-topic distribution.  Since we create each document in the corpus containing every word used in tweets related to an entity, during one day, we can retrieve the most relevant topics about an entity on a daily basis. From each of those topics we select the most related words in order to identify it. The system supports three different approaches with LDA, yielding varying results: (a) creating a single model for all entities (i.e. a single corpus), (b) creating a model for each group of entities that fit in a similar category (e.g. sports, politics) and (c) creating a single model for each entity.

\subsection{Sentiment Analysis}
A word-level sentiment analysis was made, using Sentilex-PT \cite{sentilex} - a sentiment lexicon for the portuguese language, which can be used to determine the sentiment polarity of each word, i.e. a value of -1 for negative words, 0 for neutral words and 1 for positive words. 
A visualization system was created that displays the most mentioned words for each entity/day and their respective polarity using correspondingly colored and sized circles, which are called SentiBubbles. 


\section{Visualization}

The user interface allows the user to input an entity and a time period he wants to learn about, displaying four sections. In the first one,  the most frequent terms used that day are shown inside circles (Figure 2). These circles have two properties: size and color. Size is defined by the term's frequency and the color by it's polarity,  with green being positive, red negative and blue neutral. Afterwards, it displays some example tweets with the words contained in the circles highlighted with their respective sentiment color. The user may click a circle to display tweets containing that word. A trendline is also created, displaying in a chart the number of tweets per day, throughout the two years analyzed. Finally, the main topics identified are shown, displaying the identifying set of words for each topic.

\begin{figure}[h]\label{sentibub}
    \includegraphics[height=7.0cm,keepaspectratio]{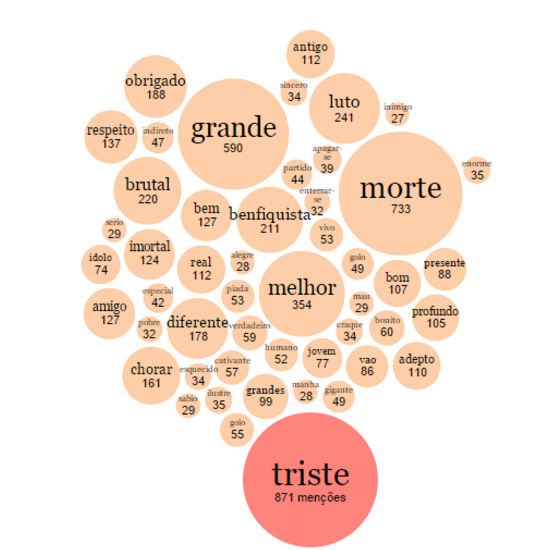}
    \caption{SentiBubbles visualization.}
\end{figure}

%
\bibliographystyle{unsrt}
\bibliography{refs}  

\begin{thebibliography}{1}

\bibitem{Saleiro2016}
Pedro Saleiro, Jorge Teixeira, Carlos Soares, and Eug{\'e}nio Oliveira.
\newblock Timemachine: Entity-centric search and visualization of news
  archives.
\newblock In {\em ECIR 2016, Padua, Italy, March 20-23}, pages 845--848.
  Springer, 2016.

\bibitem{Blei:2003:LDA:944919.944937}
David~M. Blei, Andrew~Y. Ng, and Michael~I. Jordan.
\newblock Latent dirichlet allocation.
\newblock {\em J. Mach. Learn. Res.}, 3:993--1022, 2003.

\bibitem{Kim:2004:DSO:1220355.1220555}
Soo-Min Kim and Eduard Hovy.
\newblock Determining the sentiment of opinions.
\newblock COLING '04. Association for Computational Linguistics, 2004.

\bibitem{Mei:2007:TSM:1242572.1242596}
Qiaozhu Mei, Xu~Ling, Matthew Wondra, Hang Su, and ChengXiang Zhai.
\newblock Topic sentiment mixture: Modeling facets and opinions in weblogs.
\newblock WWW '07, pages 171--180. ACM, 2007.

\bibitem{Lin:2009:JSM:1645953.1646003}
Chenghua Lin and Yulan He.
\newblock Joint sentiment/topic model for sentiment analysis.
\newblock CIKM '09, pages 375--384, New York, NY, USA, 2009. ACM.

\bibitem{saleiro2015popmine}
Pedro Saleiro, Silvio Amir, M{\'a}rio Silva, and Carlos Soares.
\newblock Popmine: Tracking political opinion on the web.
\newblock In {\em CIT/IUCC/DASC/PICOM}, pages 1521--1526. IEEE, 2015.

\bibitem{saleiro2013popstar}
Pedro Saleiro, Lu{\i}s Rei, Arian Pasquali, and Carlos Soares.
\newblock Popstar at replab 2013: Name ambiguity resolution on twitter.
\newblock In {\em CLEF 2013 Eval. Labs and Workshop Online Working Notes},
  2013.

\bibitem{sentilex}
M{\'a}rio~J Silva, Paula Carvalho, and Lu{\'\i}s Sarmento.
\newblock Building a sentiment lexicon for social judgement mining.
\newblock In {\em Computational Processing of the Portuguese Language}, pages
  218--228. Springer, 2012.

\end{thebibliography}
%
%
%
\end{document}